\documentclass[12pt, a4paper,twoside,openright]{article}

\usepackage{latexsym}
\usepackage{indentfirst}
\usepackage{graphicx}
\usepackage{amsmath,amssymb,amsfonts,amsthm,amstext,amscd}
\usepackage{mathrsfs}
\usepackage{fancyhdr}
\usepackage{times}
\usepackage{enumerate}
\usepackage{natbib}
\usepackage[hang,flushmargin]{footmisc}

\fancypagestyle{plain}{%
\fancyhf{}%
\fancyhead[LO]{\textbf{Science \& Philosophy}
}
\fancyhead[RO]{Volume 11(1), 2023 }

\fancyfoot[CO,RE]{}
}

\begin{document}
\title{\textbf{The concept of probability,\\ crisis in statistics,\\ and the unbearable lightness of Bayesing}}
\author{Boris Čulina\footnote{University of Applied Sciences Velika Gorica, Velika Gorica, Croatia; boris.culina@vvg.hr}}

\date{}

\maketitle

\begin{abstract}
\begin{normalsize}
\noindent
Education in statistics, the application of statistics in scientific research, and statistics itself as a scientific discipline are in crisis. Within science, the main cause of the crisis is the insufficiently clarified concept of probability. This article aims to separate the concept of probability which is scientifically based from other concepts that do not have this characteristic. The scientifically based concept of probability is Kolmogorov’s concept of probability models together with the conditions of their applicability. Bayesian statistics is based on the subjective concept of probability, and as such can only have a heuristic value in searching for the truth, but it cannot and must not replace the truth. The way out of the crisis should take Kolmogorov and Bayesian analysis as elements, each of which has a well-defined and limited use. Only together with qualitative analysis and other types of quantitative analysis, and combined with experiments, they can contribute to reaching correct conclusions. \footnote{\noindent Received on March 23th, 2023. Accepted on June 21th, 2023. Published on June 30th, 2023. doi: xxx. ISSN 2282-7757; eISSN 2282-7765. \copyright The Authors. This paper is published under the CC-BY licence agreement.}
\end{normalsize}
\end{abstract}

%
%

\newpage

\begin{quote}
	\textit{Probability is the most important concept in modern science, especially as nobody has the slightest notion what it means.}
\end{quote}\vspace{-8mm}
\begin{flushright}
Bertrand Russell, 1929 Lecture ‒ cited in \cite[p.~587]{bell}
\end{flushright}

\begin{quote}
	\textit{In our days, serious arguments have been made from data. Beautiful, delicate theorems have been proved, although the connection with data analysis often remains to be established. And an enormous amount of fiction has been produced, masquerading as rigorous science.}
\end{quote}\vspace{-8mm}
\begin{flushright}
	David A. Freedman  \cite[p.~35]{Free}
\end{flushright}

\begin{quote}
	\textit{“… Bayesianism may survive for a while, along with mind-body dualism, many-worlds metaphysics, and other philosophical extravagances. Meanwhile let us hope that it will be gradually discontinued in the law, medicine, seismic engineering, policy-making, and other fields where lives are at stake.}
\end{quote}\vspace{-8mm}
\begin{flushright}
	Mario Bunge  \cite[p.~173]{Bunge}
\end{flushright}

\begin{quote}
	\textit{Published research findings are sometimes refuted by subsequent evidence, with ensuing confusion and disappointment. Refutation and controversy is seen across the range of research designs, from clinical trials and traditional epidemiological studies to the most modern molecular research. There is increasing concern that in modern research, false findings may be the majority or even the vast majority of published research claims.}
\end{quote}\vspace{-8mm}
\begin{flushright}
	John P. A. Ioannidis  \cite[p.~0696]{Ion}
\end{flushright}

\begin{quote}
	\textit{It was most difficult to work when in the Republic of Croatia we had a “bus of the dead” every day, and in Split, one such bus would “arrive” every week. There was too much unnecessary dying, and such a situation throws you into depression and you keep repeating the same question to yourself}:
		
	\hspace{5mm}\textit{Is it possible that people do not understand and would rather die}
	\vspace{4mm}\hspace{5mm}\textit{than get vaccinated?}
		
		\textit{And dying from corona is awful, painful. It's not death from a heart attack where it “cuts you off” and that's it, you're gone. Here people suffer and die in terrible agony.}
	
\end{quote}\vspace{-8mm}
\begin{flushright}
	Ivo Ivić (Slobodna Dalmacija newspaper, 25. 03. 2022.)
\end{flushright}

\section{Introduction}
The above quotes are quotes from people who know what they are talking about, from the famous philosophers Bertrand Russell and Mario Bunge, through the famous statisticians David A. Freedman and John P. A. Ioannidis, to Associate Professor Ph.D. Ivo Ivić, head of the Clinic for Infectiology of KBC Split and head of the COVID Hospital in Križine. Each of the quotes testifies to one aspect of the crisis in statistics. The first quote testifies to the insufficiently clarified concept of probability on which statistics is based. The second and third quotes testify that there is a crisis in statistics itself as a scientific branch. The main source of this crisis is the conflict (in \cite{mayo}, the term "statistical wars"  is even used) of two schools of statistics, classical (frequentist) and Bayesian statistics. The theoretical basis of this crisis is a different understanding of the concept of probability. The fourth quote testifies to a true inflation of bad application of statistics in scientific research: impossibility of replication and reproduction, incorrect implementation and interpretation of statistical procedures, but also misuse of statistical tests, problem of p-values, irresponsible Bayesianism, etc.  (see e.g., \cite{Ion,Peng, Green, ASA}).  In addition to various sociological reasons, which burden the scientific community, the reason for this is insufficient understanding of statistics itself, its methods and concepts, which are based on the concept of probability. The last quote testifies that a misunderstanding of science, above all the role of statistics in science, can have profoundly serious consequences. This quote, together with the preceding one, also refers to a crisis in education, especially in statistics education. Statistics education has a significant impact on the use of statistics both in scientific research and in making decisions in everyday life. Its most important task is to teach correct understanding of the concept of probability. 

Thus, if we look at the internal reasons for the crisis (reasons that concern science itself), the main source of the crisis is in the insufficiently clarified concept of probability. The aim of this paper is to contribute to the separation of the concept of probability, which is scientifically based and verifiable, from other concepts that do not have these characteristics. The concept of probability that will be separated also gives answers to the question of when probability can help us describe and control reality and when it cannot. These answers allow us to identify the cause of the crisis in statistics: the wrong application of the concept of probability and, even worse, the application of the concept of probability where it is not applicable. These answers also give us guidelines for overcoming the crisis in statistics in all the aforementioned aspects.

In Section 2, the scientifically based concept of probability will be described. In essence, it is a mathematical probability model which, together with the conditions of its application, was established by Kolmogorov in 1933. The conditions of its application determine how probability can and cannot help us. In  Section 3 it will be described how this concept of probability affects the crisis in statistics. In  Sections 4 and 5 it will be explained why some other concepts of probability are not scientifically based. Special attention will be paid to the Bayesian concept of probability, which, although not scientifically based, has a certain heuristic value, but also carries great danger. 

Although this article covers a wide area, its main topic is the role of mathematical models and their application in the field of probability and statistics. I see the originality of the paper in 1) the explanation of Kolmogorov's concept of probability as a concept of probability based on a certain mathematical modelling of random phenomena, 2) the argumentation that Kolmogorov's concept of probability is fundamentally different from all other concepts of probability, including frequentism with which it is usually identified, 3) and the argumentation that Kolmogorov's concept of probability is the only scientifically based and verifiable concept of probability. However, the main message of the work is that Mark Twain's words “There are three kinds of lies: lies, damned lies, and statistics” returns to us today in a dramatic way: instead of significantly helping to overcome the crisis into which modern society has fallen, today statistics significantly contribute to the crisis. It is vitally important that statistics are developed and practised responsibly, and not that falsehoods are generated under the guise of science. A correct understanding of the concept of probability is crucial here.

\section{Mathematical probability models and conditions of their application}
It is usually stated that Kolmogorov formulated the axioms of probability theory in 1933 (\cite{Kolm}). The interpretation he gave to these axioms is rarely mentioned because it is viewed to only express von Mises' frequentism.\footnote{“Although Kolmogorov never abandoned his formulation of frequentism, his philosophy has not	enjoyed the enduring popularity of his axioms.”  \cite{Shaf}.} In the next section, I will argue that this view is wrong. In this section, Kolmogorov's formulation of the concept of a mathematical probability model and the conditions of its application will be briefly presented. 

It is common to consider that determinism and randomness are mutually contradictory concepts. But is it so? We say that a process takes place \textit{deterministically} if its future state is completely determined by the initial state. For example, the falling of an object from a certain height is determined by the initial conditions. If, under the same conditions, the process takes place in different ways, then we are talking about a \textit{random phenomenon}. Example: rolling a die. However, when we release a body from a certain height (always the same initial condition), if we were to measure the time of its fall with sufficient precision, we would always get a slightly different result, because the fall depends on a multitude of factors that we cannot control (air resistance, movement of the air mass, shape of the body, etc.). Thus, the randomness is also present here, although determinism is dominant. On the other hand, if when rolling a die, we could accurately determine the initial conditions of the throw, according to the laws of classical physics, the motion of the die takes place deterministically. Determinism is also present here, although randomness dominates. 

Determinism and randomness are always present, the only question is which one dominates. Thus, they are more complementary than conflicting concepts. Moreover, in science, the question of determinism and randomness arises not so much as a question of what nature is like, but more as a question of how we will describe a particular phenomenon, in a deterministic or stochastic way. That is why we will talk bellow about deterministic and stochastic (probabilistic) models of description of nature. In specific situations, a model of a certain type (or a combination of models of different types) should be selected to provide a satisfactory description of the situation. Even the basic physical theories about the world, classical physics and quantum physics, are big models of this type. One is based on determinism, the other one on randomness. According to classical physics, the world is deterministic, and randomness is the result of our inability to control all parameters of a process. According to quantum physics, in its standard interpretation, randomness is a necessary part of the description of the world: we can only determine the probabilities that something will happen.

\textit{Deterministic models} are, mathematically speaking, sets of functions that determine some quantities using other quantities. That is why the mathematics of deterministic models is in the essence the mathematics of functions. Mathematics can offer various models, and the person who applies them must determine which models are good enough for his problem. How to use mathematical models is no longer a matter of mathematics but a part of science that deals with a certain phenomenon. For example, in the case of a fall in vacuum on the Earth's surface, mathematics can offer various models of how the distance travelled \textit{s} depends on the time of fall \textit{t}: 

$$s=kt, s=kt^2,   s=kt^3, s=\dfrac{k}{t^2}, s=k\ln(at),\ldots$$ 

\noindent Physics has chosen the model that corresponds to reality: $s=  \frac{1}{2} gt^2$, where $g=9.81 ms~^{-2}$. However, this model is an approximation because g depends on the location on the Earth's surface, and we can determine g most precisely experimentally. If we analyse the real situation, the fall of a body somewhere on the Earth's surface, then we have to take air resistance into account. The previous formula is now only approximately good for heavier bodies and lower speeds. If the speeds are higher, then air resistance is more prominent, and we need a more complicated model that gives us more accurate results. However, this still remains an approximate solution. A model that would provide a more accurate solution should include airflow, temperature, humidity, pressure, vibrations, the shape of a falling body, etc. Increasingly accurate models would be increasingly complicated. But no matter how complicated they are, they could not describe the most ordinary fall of leaves in autumn under a gust of wind. The deterministic models lose their power of description because now we encounter  a random phenomenon: under the same initial conditions that we can control, the leaves will fall in different ways. Thus, to get an even more precise model, we would have to include a corresponding probability model. 

This example illustrates the general situation. The goal of any modelling is to find the simplest possible model that gives accurate enough results for our needs. When we create a model, we try to test it by comparing the values it predicts with the values we determine experimentally. If this match is good enough, that is, if from the point of view of the problem we want to solve, the obtained deviations are negligible, then we accept that model. If not, then we look for a better model, which is often more complicated. The same applies to probability models.

To describe random phenomena, mathematics offers probability models. For a simple illustration I will use the roll of an irregular traditional die. We specify a random phenomenon by specifying the conditions under which it takes place (for the die, we must precisely define the conditions under which the die is rolled). To the specified random phenomenon, we associate the outcomes space (sample space), a set whose elements we call outcomes. The only condition for the outcomes space is that exactly one outcome is associated with each realization of the random phenomenon. We can associate various outcomes spaces to the same random phenomenon depending on what we are interested in. For the die, it is usually the set $\{1,2,3,4,5,6\}$, but it can also be the set $\{even,  odd\}$, if we use the die to make a binary choice. We describe events with words that set conditions on outcomes (e.g., that a number divisible by 3 will be rolled), and “officially” define events as sets of outcomes that satisfy a given condition (divisibility by 3 gives the set $\{3,6\}$). We cannot predict what will happen in a single roll of the die. However, a regularity emerges in the sequences of rolls. In each sequence of rolls, we can associate a relative frequency $\dfrac{n_D}{n}$ to the event $ D $, where  $ n_D $ is the number of rolls in which $ D $ occurred, and $ n $ is the total number of rolls in the given sequence. It is an experimental fact that for each such sequence of rolls the relative frequencies are approximately the same. The higher the number of rolls, the better the grouping of the frequencies. The number around which the relative frequencies are grouped is called the probability of the event $ D $ and denoted by $ P(D) $. \textit{That number is not uniquely determined by the grouping, but we postulate it}. The meaning of this number is that it approximates (estimates) the relative frequency of the event \textit{D} in a sequence of rolls: 

$$P\left(D\right)\simeq\frac{n_D}{n}$$

\noindent It follows from this meaning that the probability function \textit{P}, which associates a probability with each event, cannot be an arbitrary function. Given that it approximates relative frequencies, it must have properties of relative frequencies for a given sequence of rolls. Such are the following properties, which we call probability axioms, because other properties can be derived from them:\footnote{These axioms are not independent: the third property and one of the properties under 2. can be derived from the other properties.}

\begin{enumerate}
	\item 	$ 0\le P\left(D\right)\le 1 $
	\item  $ P\left(\emptyset\right)=0 $, $ P\left(\mathrm{\Omega}\right)=1 $
	\item  $ A\subseteq B\ \longrightarrow P\left(A\right)\le P\left(B\right) $
	\item  For mutually exclusive events $ A $ and $ B $, $P\left(A\cup  B\right)=P\left(A\right)+P\left(B\right) $
\end{enumerate}

\noindent where $ \mathrm{\Omega} $ is the outcomes space. 

By modelling the random phenomenon of rolling an irregular die, we arrived at the (almost) general concept of Kolmogorov's probability model of a random phenomenon. A \textit{finite probability model} consists of:

\begin{itemize}
	\item 	a non-empty finite set $ \mathrm{\Omega}\ $ which we call the \textit{outcomes space}, its elements \textit{outcomes}, and its subsets \textit{events},
	\item and the function \textit{P}, which we call \textit{probability}, which assigns a number to each event, whereby the above axioms of probability must be fulfilled.
	
\end{itemize}

\noindent Here we limited ourselves to finite models because infinite models (e.g., random selection of a point from a geometric figure) have somewhat more complicated mathematics, while the basic ideas are the same as for finite models.

As with deterministic models, mathematics offers various probability models too. By choosing a non-empty set and a function that assigns a real number to each subset, where the function satisfies the stated axioms, we obtained a probability model. The main problem is to choose a model for a given random phenomenon that describes the phenomenon well. The basic criterion for the correct choice of a probability model, set by Kolmogorov in 1933 (\cite[p.~5-6]{Kolm}), is that the probability $ P(D) $ of an event $ D $ must match well enough the relative frequencies of the event in a sequence of repetitions of the random phenomenon:

\begin{center}
	$P\left(D\right)\simeq\dfrac{n_D}{n}$ \hspace{1cm} (1)
\end{center}

 \noindent What is “well enough”? The answer is the same as with deterministic models: from the point of view of the problem we want to solve, the deviations must be negligible. This approximate equality is the basic connection between a probability model and reality, the basic criterion for the usability of the model in the description of a given random phenomenon. Thanks to this connection, the use of probability models is as firmly scientifically based as the use of deterministic models. On the one hand, if we know from some theory or experiment the probability of an event, then by this connection we can estimate the relative frequency of its occurrence in a series of repetitions of a given random phenomenon. On the other hand, if we determine the relative frequencies of events through experimentation, then by this connection we can estimate their probabilities. If we analyse how probability is used in classical statistics, we will see that it is precisely this concept of probability.

Let us not forget that the basic assumption of this modelling is that a random phenomenon can be replicated an arbitrary number of times. Thus, Kolmogorov's overall conception sets two criteria for the usability of probability models:

\begin{itemize}
	\item 	the basic condition for applicability of probability models: \textit{probability models can be applied to random phenomena that can be replicated an arbitrary number of times}.
	\item 	the basic condition for the correct choice of a probability model for a given random phenomenon: \textit{ the probability of an event must match well enough with the relative frequencies of the event in a sequence of repetitions of the random phenomenon.}
	
\end{itemize}

\noindent When the stated conditions are met, probability models successfully describe what will happen in a sequence of repetitions of a random phenomenon. This is confirmed by the exceptional success of these models in statistical mechanics and quantum physics, but also in industry, including the “industry” of insurance companies and games of chance. By taking a random sample from a population, the success of these models is extended to biology, medicine, and social sciences. By ensuring that each individual has the same probability of being selected into the sample, the population parameters become the parameters of the probability model of random selection of an individual from the population. Then the basic connection (1) relates the relative frequencies in the entire population to the relative frequencies in the sample. However, the problem of obtaining a random sample arises here. The deviation of the actual sample from a random sample can drastically devalue the obtained results.

The basic connection (1) between a probability model and reality also tells us when probability cannot help us in describing,  predicting, and controlling reality. A typical such situation is a single realization of a random phenomenon (e.g., which number will fall in one roll of a die). In a particular case, an event will happen or not happen. \textit{Except in some extreme situations, probability cannot tell us whether an event will happen or not}. It is the so-called \textit{problem of the single case}. Extreme situations are when the probability of an event is close to 0 or 1 to a few hundredths or smaller. If the model is good, in the first case we can \textit{practically be sure} (Kolmogorov's term) that the event will not happen, while in the second case we can practically be sure that it will happen. Likewise, if the probabilities of the event are close to those extreme values, and we must make a decision, then it is more reasonable to make the decision that the event will not happen (if the probability is close to 0) or that it will happen (if the probability is close to 1), although we do not have absolute certainty.

Take, for example, the problem of vaccination, which dominated the Covid-19 pandemic. Roughly speaking, the data show that the probability that someone will die from vaccination is at most in the order of millionths, and that they will die from Covid-19 (unvaccinated, because the vaccinated are practically certainly protected) during a pandemic is in the order of thousandths. According to this rough but basically correct model, in a larger population, such as a population of 4 million people, roughly the size of Croatia, if everyone chooses not to vaccinate, about $ 4000 $ will die, and if everyone chooses to vaccinate, about 4 will die. Thus, at the level of the whole society (population), it is correct to recommend vaccination. This will save thousands of lives. But what about the individual? A choice that is good for the whole society is not necessarily good for an individual. Although both probabilities are exceedingly small, the probability that an individual will die from vaccination is  a thousand times smaller than they will die from Covid-19, and it is reasonable, if we have no other knowledge, to choose vaccination. Although probabilities cannot deprive us of uncertainty in single cases, they are still an important indicator that tells us which decision is more reasonable to make.

The above probability model of the Covid-19 pandemic has another significant drawback related to the probability it assigns to the occurrence of an event in a particular case. This is the so-called \textit{problem of the referent random phenomenon}: \textit{the probability of an event depends not only on that event, but also on the random phenomenon in relation to which we observe it.} The model assigned the probability of death from Covid-19 to an individual in relation to the entire population. If we narrow the population to which that person belongs by setting additional conditions (age, specific health problems, ...) then this probability changes and increasingly suits that person. For example, the following graph (Figure 1) shows what percentage of people in each age group in Croatia died from Covid-19.\footnote{The graph is derived from the data on the 2021 population census of the The Croatian Bureau of Statistics and data on mortality from COVID-19 from the Croatian Institute of Public Health.}

\begin{center}
	\includegraphics[scale = 0.9]{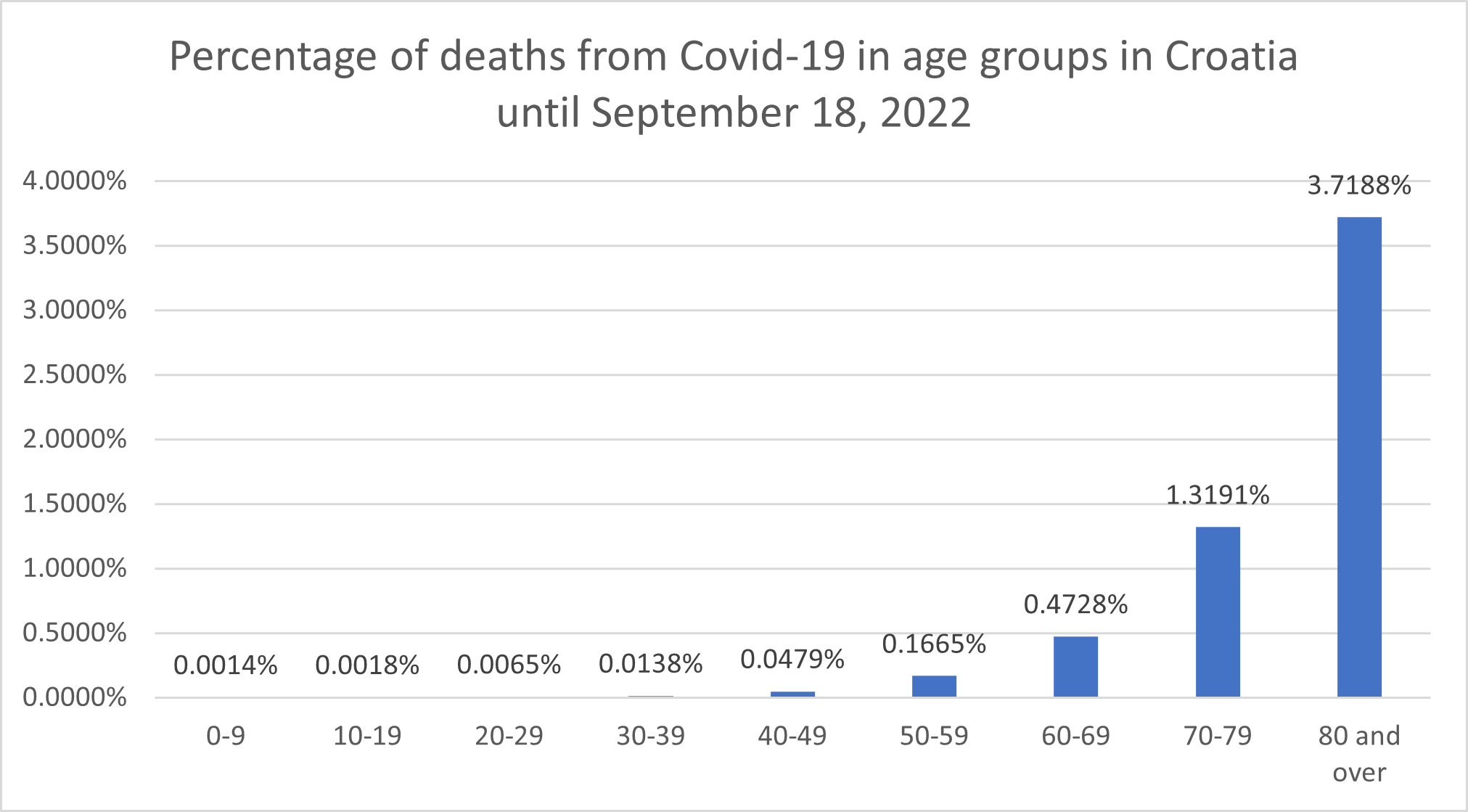}
	
	Figure1: Mortality by age group
\end{center}

\noindent These percentages estimate the probabilities that someone in each age group will die from Covid-19. In relation to the probability of death from Covid-19, which is about one thousandth for the entire population, for the population of people over 70 years old the probability is drastically higher, about one hundredth, and for those under 30 years drastically lower, about one hundredth thousandths. For an individual, the probability for his age group better describes his chance of dying from Covid-19 than the probability for the entire population. This dependence of the probability of an event on the population in relation to which we observe it has no effect on the estimates of the relative frequencies of the event in the populations, but it influences the significance of the probability for an individual.

Despite these shortcomings of the concept of probability related to predicting what will happen in an individual case – that probability makes predictions only for the population and not for the individual, and that this probability depends on the population to which the individual belongs – we can still apply this understanding of the concept of probability to the decision to vaccinate. Ideally, for each individual, we could determine such a narrow reference population to which they belong, and for which the probability of death from vaccination for a person from that group is equal to 1 (everyone from that group will die if vaccinated) or 0 (everyone from that group will not die if vaccinated). If science could find such a narrow group for each person, then we would know exactly who can and who cannot receive the vaccine. However, too many parameters are involved, and the processes are too complex for us to have full understanding and control over the outcomes. It can be said that the probability models we currently have are also measures of our knowledge: the more specific the models, the greater the knowledge. In the short time it had at its disposal, science managed to 1) determine the probabilities of death for not particularly specific groups (the simple model  described above is the crudest such model), 2) give general instructions about behaviour, and 3) identify specific groups that could be extremely vulnerable to vaccination or to the COVID-19 disease. Thus, science has given a clear instruction: if the person does not belong to an extremely vulnerable group for vaccination, they should be vaccinated. 

\section{The concept of probability and crisis in statistics}

The scientifically based concept of probability described above is highly successful within the limits of its applicability. However, these limits are very strict and probability models cannot be applied to many situations to which we would like to apply them. Real random phenomena only approximately correspond to random phenomena that we can always reproduce under the same conditions. The samples we use to make inferences about a population often deviate too much from random samples, preventing credible conclusions to be drawn. Furthermore, in more complex situations we would need many assumptions to associate a probability model with a random phenomenon. If the assumptions are not met then we have no guarantee that the predictions based on the model are correct. When it comes to an individual case, probability models lose their power and the predictions they provide can only be one of the indicators to guide us. 

The problems of complexity and of the single case of a random phenomenon are especially important for disaster risk assessments because no matter how similar they may appear, disasters are always complex and unique phenomena. In the article \cite{Free1}  it was shown how powerless statistics is in such cases. The article begins with the following text:

\begin{quote}
	What is the chance that an earthquake of magnitude 6.7 or greater will occur before the year 2030 in the San Francisco Bay Area? The U.S. Geological Survey estimated the chance to be $  0.7 \pm 0.1 $ (USGS, 1999). In this chapter, we try to interpret such probabilities.
\end{quote}

and ends with the following words:

\begin{quote}
Another large earthquake in the San Francisco Bay Area is inevitable, and imminent in geologic time. Probabilities are a distraction. Instead of making forecasts, the USGS could help to improve building codes and to plan the government’s response to the next large earthquake. Bay Area residents should take reasonable precautions, including bracing and bolting their homes as well as securing water heaters, bookcases, and other heavy objects. They should keep first aid supplies, water, and food on hand. They should largely ignore the USGS probability forecast.
\end{quote}

\noindent After a detailed analysis of how the U.S. Geological Survey came to this probability estimate, Freedman and Stark showed that due to the complexity of the mechanism of the earthquake, many assumptions and subjective assessments had to be made, which completely devalued the established model and the obtained probability estimate.

David A. Freedman is one of the well-known statisticians who focused on the issue of the correct use of statistics, pointing out today's wrong practice of statistics where statistics are used “everywhere” and “unlimited” in such a way that “without much trouble” researchers arrive at mostly wrong answers. In contrast, Freedman advocates a balanced combination of qualitative and quantitative research, rather than blindly accepting statistical inferences based on untested assumptions.\footnote{“But perhaps the most serious source of error lies in letting statistical procedures make decisions for you.”  \cite[p.~3]{Good}.} Statistical models and methods should be formulated carefully: their assumptions should be explained, their validity evaluated, and experimentally verified on several samples. The results obtained in this way should be viewed through qualitative analysis and research in order to finally obtain credible conclusions.\footnote{“Combining qualitative insights and quantitative analysis - and a healthy dose of scepticism - may provide the most secure results.” \cite[p.~313]{Free2}.} Freedman called this approach the \textit{shoe leather approach}.

The above consideration shows that Kolmogorov's models have strict limits to their applicability, and even where they can be applied, they often require a shoe leather approach. This makes classical statistics a “hard” science with limited possibilities. And this is precisely where the cause of the crisis in statistics and its application lies: in the conflict between these characteristics of statistics and the needs and desires of modern society to solve as many problems as possible and in as little time as possible with statistical methods. The conflict occurs in two segments: where statistics can be applied and where statistics cannot be applied in a scientifically objective manner.

Where statistics can be applied in a scientifically objective way, in the analysis of mass phenomena, the applications are not done with sufficient quality, either because of ignorance and irresponsibility, but also because of unethical behaviour.\footnote{For example, a recent comprehensive survey of science in the Netherlands  \cite{Dutch} showed that in the last three years approximately every second scientist participated in at least one qualitatively questionable research, while every twelfth scientist faked or fabricated their research at least once.  }  That is why today the scientific community places more and more demands on financiers and academic institutions to ensure an adequate system that would not encourage unethical activity\footnote{Unethical activity is largely encouraged by the existing system of financing and evaluation of scientists, which does not respect equality and diversity in science and sets narrow bureaucratic evaluation systems. This has already become such a big problem that there is a distinct trend in the scientific community to completely change the system of doing science and its evaluation. }, and would support more rigorous scientific research: “(1) more courses on methodologically robust and transparent scientific research in student curricula and training of established researchers; (2) increase in the involvement of experienced methodologists, statisticians and data stewards on projects by, for example, securing funding for such personnel or establishing advisory bodies that would provide advice and guidance for funded projects; (3) better technical/infrastructural support for enabling open science practices, rigorous reporting, archival of all elements of research and creating linkages among them” \cite[p.~1396]{Aca}.

Far more worrying is the ubiquitous use of statistics where its scientific power is very weak — in predicting what will happen in a particular case under conditions of uncertainty. Kolmogorov's models can give the probability of whether something will happen in a particular case only if that case is the realization of a well-defined and repeatable random phenomenon. But even then, those probabilities do not mean much. Only when these probabilities are close enough to one or zero can we practically be sure that something will happen or not happen. Otherwise, it is reasonable to stick to probabilities in predicting only when we do not have better deterministic indicators. However, uncertain situations are frequent and very important, and require our reaction, especially if they are crisis situations. This creates a strong need to try to provide predictions as a basis for appropriate action in such situations. Since classical statistics is almost unusable there, there was an accelerated development of Bayesian statistics, as an alternative to classical (frequentist) statistics. Bayesian approach is based on a completely different concept of probability, the so-called concept of subjective probability, which will be briefly explained below. Such a concept of probability allows Bayesian statistics to be applied everywhere. That is why to many researchers Bayesian statistics seems to be the “holy grail” of statistics which solves all problems almost algorithmically. The danger lies in the fact that the truth about Bayesian statistics is exactly the opposite: its meaningful use requires an even stronger shoe leather approach compared to frequentist statistics.  The danger that Bayesian statistics carries is vividly stated in \cite[p.~447]{Gel8}: “\ldots recommending that scientists use Bayes’ theorem is like giving the neighbourhood kids the key to your F-16”.  It should be kept in mind that almost everyone except very experienced statisticians are “neighbourhood kids” here, and that the “F-16” flies through vital areas of human society: from politics, judiciary, economy and finance to health, environmental protection and natural and social threats.

Unfortunately, the crisis is not only in inappropriate application of statistics. In statistical science itself, there has been a crisis due to the division among statisticians regarding the preference for classical or Bayesian statistics and criticizing the other.\footnote{“Basically, there's only one way of doing physics, but there seems to be at least two way of doing statistics, and they don't always give the same answers.” \cite[p.~1]{Ef}.}  Of course, this crisis also produces additional confusion in the application of statistics. The situation becomes even more dramatic with the advent of \textit{Big Data}, which poses new major challenges for statistics.\footnote{“Perhaps Statistics should stand up for its responsibilities before a Big Data Disaster” \cite[p.~7]{Fra}.} The statistical community is aware that the time has come to overcome the crisis.\footnote{“This will be a challenging period for statisticians, both applied and theoretical, but it also opens the opportunity for a new golden age, rivalling that of Fisher, Neyman, and the other giants of the early 1900s.” \cite[p.~1]{Ef}.} Understanding the various concepts of probability is crucial here, especially Kolmogorov's concept of probability and Bayesian concept of subjective probability. In the next section I will argue that Kolmogorov's interpretation differs significantly from the frequentist and propensity interpretation. Thus, it does not fall under the criticism that these other two interpretations fall under. In Section 5, I will highlight some essential characteristics of Bayesian interpretation that determine its place in statistics. The correct understanding of Kolmogorov’s and Bayes’ interpretation indicates the way out of the crisis.

\section{Kolmogorov's interpretation versus frequentist and propensity interpretation}

Kolmogorov's interpretation of probability is usually placed in the frequentist interpretation of probability elaborated by von Mises in 1919 (\cite{Mises}). However, one “detail” significantly separates Kolmogorov's interpretation from the standard von Mises frequentist interpretation. Von Mises postulated that the relative frequencies $\dfrac{n_D}{n}$ of an event \textit{D} in a series of repetitions of a random phenomenon converge to a certain number in the sense  of the mathematical concept of limit and that this very number is the probability of the given event:

$$P\left(D\right)=\lim_{n\rightarrow\infty}{\frac{n_D}{n}}$$

\noindent This postulate is experimentally unfounded because neither an infinite number of repetitions of a random phenomenon is possible nor experimentally obtained finite sequences of relative frequencies support such a claim. Criticisms of the frequentist interpretation are mainly criticisms of this postulate. This postulate is simply “too strong”. In addition, this postulate gives the concept of probability a completely objective meaning in the sense that probability belongs to a random phenomenon in the same way as, for example, the number of elements of a set belongs to that set. In contrast to such an approach, Kolmogorov's interpretation of the concept of probability has a completely different basis. It rests on the idea of modelling: probabilities are ideal numbers within a probabilistic model that approximate the relative frequencies well enough:

\begin{center}
	$P\left(D\right)\simeq\dfrac{n_D}{n}$ \hspace{1cm} (1)
\end{center}

Using (1), we can experimentally set up a probabilistic model for the given random phenomena, as well as check whether the probabilistic model is good enough for the given random phenomena. And this is enough -- Kolmogorov's interpretation does not go beyond this scientifically based and verifiable setting of probabilistic models of random phenomena. In this interpretation, the probabilities are as objective as the above connection is.\footnote{I would point out that most other scientific concepts have the same kind of objectivity as probability. For example, we associate the length s of the travelled path with free fall. However, s is the distance between two points in Euclid's ideal geometric model of space. This number is as objective for free fall as the connection of two points in Euclidean space with the beginning and the end of the fall is objective. In macroscopic conditions, this connection is strong, but in microscopic conditions, instead of fixed points, we have molecules in constant oscillation, so we have an indeterminacy in what is the initial and what is the end point of the fall. This connection becomes even looser if we go to even smaller dimensions where the laws of quantum physics dominate. Thus, like probability, the length of the travelled path in free fall is an ideal element of the corresponding physical model which is as objective as the concept of the travelled path is objective.} Although von Mises' and Kolmogorov's formulas for $ P(D) $ seem to be remarkably similar, their interpretive basis, and thus the concept of probability, are fundamentally different.\footnote{Kolmogorov's interpretation is often mistakenly identified with the so-called finite frequentism, according to which (1) is taken as the operational (empirical) definition of probability. A probability model is a much more subtle concept than the operational “definition” of probability, so the corresponding interpretations of probability differ significantly.}

Likewise, the propensity theory, whose most famous representative is Karl Popper (see e.g., \cite{Pop}), at first glance seems to be only an addition to Kolmogorov's interpretation in the part that refers to the individual realization of a random phenomenon. However, this addition leads to a completely different understanding of probability. Namely, Kolmogorov's models do not say anything about an individual realization, but only predict ratios in a sequence of such realizations. This is paradoxical at first sight - the sequence consists of individual realizations, and how can regularity appear in the sequence if it is not present in the individual realization? Popper introduces the term propensity (tendency) to give probability a meaning for a particular realization of a random phenomenon. According to Popper, the conditions that determine random phenomenon are “endowed with a tendency, or disposition, or propensity, to produce sequences whose frequencies are equal to the probabilities” \cite[p.~35]{Pop}. According to Popper, propensities are present in every realization of a random phenomenon, just as, for example, the tendency to fight for survival is present in living beings. Thus, probabilities as measures of these propensities are also present in every realization of a random phenomenon. In a series of realizations, these propensities are manifested as tendencies to obtain relative frequencies equal to the stated probabilities. This addition to Kolmogorov's interpretation of probability, like frequentism, introduces the concept (propensity) which is experimentally unverifiable, and which gives an objective interpretation to the concept of probability. Thus, the propensity interpretation, as well as frequentism, is not a completion of Kolmogorov's interpretation, but it cancels its basic philosophy derived from the modelling of random phenomena and replaces it with a completely different concept of probability.\footnote{In addition to unverifiability, the concept of propensity has another important drawback: it is in contradiction with the problem of referential random phenomena. The probability of an event depends on the conditions by which we determine the random phenomena. Under one set of conditions, a random process will have one probability, and under another a different probability. Thus, propensity cannot be an intrinsic characteristic of a particular process, but a reflection of its relationship to the conditions under which we determine the random phenomena, which often depend on our modelling. }

\section{Subjective interpretation, Bayesian statistics and Bayeseanism}

In contrast to Kolmogorov's interpretation of probability, which through the formula $P\left(D\right)\simeq\dfrac{n_D}{n}$ has its objectivity in the sense of checking whether it corresponds well enough to reality, Bayesian statistics rests on a completely different concept of probability according to which probability is a subjective measure of our belief in the truth of a statement. We use this interpretation in everyday life, for example when we say that it is more likely that there is life outside our planet than that there is not. If we are talking about whether an event will happen, for example when I say that I am almost 80\% sure that it will snow tomorrow, then the stated percentage is a subjective measure of my belief that the event will happen. The concept of probability that dominated until the beginning of the 20th century belongs to the subjective concept of probability: probability is a measure of our ignorance. Such an interpretation is a product of an era in which, under the influence of Newtonian mechanics, it was believed that processes in nature take place deterministically.\footnote{The history of the concept of probability is nicely presented in the book \cite{Plato}.} Frequentist conception of probability was affirmed in the first half of the twentieth century, primarily due to its successful application in statistical and quantum physics, as well as in classical statistics. The subjective conception of probability appears in a new form through the works of de Finetti\footnote{See for example \cite{Fin}} and Savage\footnote{See for example \cite{Save}}, no longer as a measure of our ignorance, but as a measure of our belief in the truth of various claims. The axioms of probability now have the role of rational conditions on the subjective assignment of probability. For example, axiom $A\subseteq B\ \longrightarrow P\left(A\right)\le P\left(B\right)$ now has the following interpretation. Let $ A $ and $ B $ be two statements such that if $ A $ is true then $ B $ is also true. Then the statement $ B $ must have a greater or equal subjective probability to be true than does the statement $ A $. When, based on our knowledge or experience, we use probabilities to determine how much we believe some statements to be true, the calculus of probabilities allows us to consistently calculate the probabilities that some other statements are true. Specially, if we are dealing with predictions in a random phenomenon, this calculus allows us to calculate the subjective probabilities that various events will happen.

A whole series of questions are raised, to which this concept of subjective probability has not yet given a  satisfactory answer: 1) Can we measure our beliefs with numbers at all? 2) Are these numbers uniquely determined by our beliefs? 3) How correct is it to assume that these numbers satisfy Kolmogorov's axioms, i.e., to what extent is Kolmogorov's probability calculus a rational calculus of our beliefs?\footnote{In \cite{Free}, one can find arguments that give negative answers to these questions.} However, the most important of all is the question of what meaning the subjective probability that a statement is true has. For example, we may have various beliefs about person X, the potential perpetrator of the crime we are investigating. If we can express these beliefs with some numbers, subjective probabilities, and if we can apply to them a probability calculus that will calculate the subjective probability that X is the perpetrator of the crime, what does this subjective probability give us? He did or did not commit a crime, and the calculated subjective probability of claiming that he committed a crime is at best a reflection of our subjective attitude towards it. A more correct way is the one used in trials: to complete the argumentation (evidence) to the point where we are sure of his guilt “beyond all reasonable doubt”. 

The problems presented above also refer to the subjective probability of an event in a random phenomenon. We have seen that Kolmogorov's probability cannot predict whether an event will occur in a particular realization of a random phenomenon, except in extreme situations when the probability of the event is close to 1 or 0, and when the phenomenon is replicative. However, it can give predictions for a sequence of realizations of such a phenomenon. Subjective probability cannot predict what will happen in a particular case, even when the subjective probability is close to 1 or 0, nor what will happen in a sequence of repetitions of a random phenomenon. Thus, it is inferior to the Kolmogorov probability in predicting the occurrence of an event in a sequence of repetitions of a random phenomenon. In predicting whether an event will occur in a particular realization of a random phenomenon, both interpretations are weak. However, the Kolmogorov probability retains at least some objectivity based on objective knowledge of what happens over a sequence of repetitions and almost certainly gives a correct prediction if the probability is close to 1 or 0. Subjective probability only expresses our attitude towards the occurrence of that event and is not an objective measure of that occurrence, no matter how we calculated it. However, as a subjective measure, it can have a significant impact on our further action.

The calculus of subjective probability is often presented as the logic of reasoning. Classical logic deals with the truth of declarative statements. If in classical logic we have proved a statement $ X $ from some assumptions, then we are sure that if the assumptions are true in a certain situation, the statement $ X $ will also be true. This has an objective meaning. If we experimentally established that the assumptions are true, we know that $ X $ is also true. If we experimentally established that $ X $ is not true, then at least some of the assumptions are also not true. Just as classical logic allows us to calculate the truth of the statement $ X $ from the truth of the assumptions, so the calculus of subjective probability  allows us to calculate the probability of the statement $ X $ from the probability of some other statements. However, there is no objective verification here because the subjective probabilities do not say anything about the truth of the statements, but only about our beliefs in their truth.

It follows from the above that the subjective concept of probability is “tempting” because, unlike Kolmogorov's probability, we can apply it everywhere. However, it is not scientifically based because it cannot predict whether a statement is true or not, that is, whether an event will happen or not, or even what its relative frequency is in a sequence of repetitions. It only has a heuristic value that can only subjectively influence our further actions. But even then, even if an expert determines subjective probabilities for a certain area, they must be only one set of parameters used to guide further action. When it comes to situations in which wrong actions can have profound consequences, then additional research should be done that will almost determine what will happen, and decisions should be made whose correctness we are sure of, based on the acquired knowledge, and not relying on subjective probability, as clearly argued in \cite{Bunge}. In short, Kolmogorov's probability is a successful result of mathematical modelling of repeatable random phenomena, while subjective probability is a questionably successful result of mathematical modelling of our beliefs in the truth of statements, and this determines the limits and way of applying these concepts.

The subjective interpretation of probability is the basis of Bayesian statistics. Thus, Bayesian statistics inherits the characteristics of subjective probability. This basis enables it to be applied everywhere, but it also determines that its results have only heuristic and not scientifically based value. The specificity of Bayesian statistics is that it introduces a certain dynamic mechanism on this basis, which has its driver in  Bayes' formula. 

To state  Bayes' formula, we need the notion of conditional probability. For each event $ A $ that has a non-zero probability, we can define a new function $P(\quad|A)$ on the events space that assigns a number $P(X|A)$ to each event $ X $:

$$P\left(X\middle|\ A\right)=\frac{P\left(X\cap A\right)}{P\left(A\right)}$$

\noindent This function also satisfies the probability axioms and is called the \textit{conditional probability}. In Kolmogorov's interpretation, in a sequence of repetitions we only look at how many times $n_{X\cap A}$ an event $ X $ happened in those $n_A$ occurrences when an event $ A $ happened. The conditional probability corresponds to the relative frequency 

$$P\left(X\middle|\ A\right)\simeq\frac{n_{X\cap A}}{n_A}$$

\noindent In Kolmogorov's interpretation, this connection gives the conditional probability the same objectivity as the probabilities of the original probability model. In the subjective interpretation, $P\left(X\middle| A\right)$ determines how confident we are that statement $ X $ is true if statement $ A $ is true.

\textit{Bayes' formula} follows from the probability axioms:

$$P\left(H\middle|\ A\right)=\frac{P\left(H\right)}{P\left(A\right)}P\left(A\middle|\ H\right)$$

In Kolmogorov's interpretation, Bayes' formula has a precisely determined objective meaning. In a sequence of repetitions, it connects the relative frequency of an event $ A $ under the condition that an event $ H $ has occurred with the relative frequency of the event $ H $ under the condition that the event $ A $ has occurred. The importance of Bayes' formula is that we can determine the probability of an event in a narrower group from knowing the probability of an event in a wider group. And we have seen how important narrowing of groups (increasing the conditions for a random occurrence) is for predicting what will happen to an individual from the group (what will happen in a particular realization of a random occurrence). As an illustration, let us take the probability of death from vaccination against Covid-19 for a given person. Let $ H $ be the event that a person from a certain population will die from vaccination, and let $ A $ be that he has certain physical characteristics (vulnerability to certain allergies, advanced age, weak immunity, etc.). If we know for the entire population the probability of death from vaccination $ P(H) $, the probability $ P(A) $ of physical characteristics $ A $ and the probability $ P\left(A\middle| H\right) $ that a person had these characteristics if he died from vaccination (all of the above can be experimentally estimated via relative frequencies), then Bayes' formula gives the probability $ P\left(H\middle| A\right) $ of death from vaccination for a narrowed group, the part of the population that has characteristics $ A $. By finding additional characteristics of a given person, we can again apply Bayes' formula to find the probability of death from vaccination for a given person  compared to an even narrower group. In this way, we not only get probabilities for smaller groups, but we also update probabilities for an individual (a single realization of a random phenomenon). If we thereby get probabilities closer and closer to 1 or 0, then we can practically determine with certainty whether a person will die from vaccination or not. This is the real power of the Bayes' formula (and in general, Bayesian analysis on Bayesian networks).

However, the unlimited application of this formula in the subjective interpretation of probability is precisely what characterizes the philosophical doctrine of \textit{Bayesianism}, while its application in the subjective interpretation of probability to statistics characterizes \textit{Bayesian statistics}. It should be noted right away that it is questionable whether the Bayes' formula can be applied to statements at all because it is derived from Kolmogorov's axioms, for which it is also questionable whether and under what conditions they can be applied to statements. In subjective interpretation $ H $ is usually a hypothesis while $ A $ is a verifiable fact. Specifically, in Bayesian statistics, $ H $ is a hypothesis about some parameter of a random phenomenon or about a model of a random phenomenon, while $  $A is a record from a sample of that random phenomenon or population. Then $ P(H) $ is the so-called a priori subjective probability that the hypothesis $ H $ is true, while $ P\left(H\middle| A\right) $ is the posterior updated subjective probability that $ H $ is true after it has been recorded that $ A $ is true. In this way, Bayes’ formula allows us to update the subjective probability of the truth of $ H $ with each new knowledge. However, a new problem arises here: it turns out that this updating is highly dependent on the initial subjective probability. Or in the words of D. A. Freedman \cite[p.~xiv]{Free3}: “to pull a rabbit from a hat, a rabbit must first be placed in the hat.” Moreover, it can be seen from the Bayes' formula that with an appropriate choice of a prior probability we can obtain any desired a posterior probability, so “…any reaction to any experiences can be justified in advance. Thus, Bayesianism implies ‘anything goes’\ldots”  \cite[p.~114]{Alb}. That is why not only classical statisticians but also Bayesian statisticians warn of the dangerous consequences of irresponsible use of Bayesian statistics. Even when an expert of a field determines the initial probabilities according to his beliefs and revises these probabilities based on new records using the Bayes formula, all the obtained results remain closed in the subjective relationship of the researcher to the phenomenon under investigation. The results have no objective weight, but only a heuristic weight the only meaning is that it can guide the researcher in further research using other methods. 

Making conclusions only based on Bayesian analysis with subjective probabilities, and not yet supported by the experience of researchers but by black box algorithms that run on computers for which the user often does not know what exactly they are doing, is simply the generation of nonsense that is sold for correct conclusions. The unbearable lightness of Bayesing consists precisely in the fact that such “science” can be done by anyone in any field, which is very dangerous. That is why Bunge   writes: “…if the U.S. Food and Drug Administration and the various agricultural experimental and forestry stations were ever to be dominated by Bayesians, they would become public perils rather than guardians of public health…” \cite[p.~173]{Bunge}, and “subjectivism … is a mark of either anti science or pseudoscience. Learned ignorance is still ignorance. In particular, diagnosis with the help of made-up numbers is not safer than without them; and trial by numbers is no more fair than trial by water or by combat. Thou shalt not gamble with life, justice, peace, or truth.” \cite[p.~173]{Bunge}. A recent disastrous example of Bayesianism is the scandal in the Netherlands, when between 2013 and 2018, the tax authorities wrongly accused tens of thousands of parents of fraudulently obtaining tax benefits. Some parents had to repay tens of thousands of euros to the state, which pushed many families into a financial and psychological crisis: from loss of housing to divorce. It all started when, in 2013, the tax authority introduced a fraud detection algorithm based on the so-called “risk classification model” that performed a Bayesian analysis with subjective probabilities that included, among other things, racist and ethnic prejudices (\cite{Xen}).

The heuristic value of Bayesian analysis with subjective probabilities, to guide us in the search for the correct theoretical model, also has its limits. Hume already showed that we cannot discover a theoretical model (universal laws) by any form of induction based on evidence (Hume's problem of induction), and the Bayesian mechanism tries to do just that.\footnote{“Bayesian rationality is just a complicated restatement of Hume’s irrationalism” \cite[p.~107]{Alb}} Furthermore, Pierre Duhem (\cite{Duh}) showed already at the beginning of the twentieth century, and Willard Van Orman Quine (\cite{QuineT}) later generalized it to a wider context (the so-called Duhem–Quine thesis) that every theoretical model must be seen as a whole. If the model incorrectly predicted a result, then it basically fails as a whole, and that fail cannot be attributed to just one hypothesis in the model. Usually this is followed by “repairing” the model and in the final repair one part is changed, maybe only one hypothesis. Bayesian analysis reduces updating the model to a single hypothesis in advance, and is therefore limited. 

The responsible use of Bayesian analysis with subjective probabilities and Bayesian statistics derives from its heuristic value. As with Kolmogorov's models, if we know the limits of applicability of Bayesian analysis, how it can help us and how it cannot, it can be very useful. It should only be clear that due to its subjective nature, it can only play a heuristic role in the search for a theoretical model of a phenomenon, and when it comes to a random phenomenon ‒ the corresponding Kolmogorov model. Combined with other considerations, it can help us find the correct model. But the final test of the correctness of the model is the experimental test ‒ confronting the model with the reality it is trying to describe ‒ and here there is no more room for Bayesian analysis with subjective probabilities. 

\section{Conclusion}

Kolmogorov's concept of probability is the only scientifically based concept of probability. It has clear conditions of its applicability and forms the basis of classical statistics. Bayesian statistics is based on the subjective concept of probability, and as such can only have a heuristic value in searching for the truth, but it cannot and must not replace the truth. The following quote succinctly illustrates the different role of classical and Bayesian statistics: “The FDA , for example, doesn’t care about Pfizer’s prior opinion of how well it’s new drug will work, it wants objective proof. Pfizer, on the other hand, may care very much about its own opinions in planning future drug development.”  (\cite[p.~1]{Ef}). The way out of the crisis and a new synthesis of statistics as a scientific discipline should take Kolmogorov and Bayesian analysis as elements, each of which has a well-defined and limited use. Only together with qualitative analysis and other types of quantitative analysis, and combined with experiments, they can contribute to reaching correct conclusions. (see e.g.  \cite{Gel13} and  \cite{Little}).

\bibliographystyle{abbrvnat}
\bibliography{Probability}

\end{document}